\documentclass[aps,prl,twocolumn,showpacs]{revtex4}

\usepackage{epsfig}
\usepackage{graphicx}
\usepackage{amsmath}
\usepackage{bm}

\begin{document}
\setlength{\columnwidth}{0.7\textwidth}

\title{Long-range ferromagnetic correlations between Anderson impurities
in a semiconductor host}

\author{N. Bulut$^{1,2}$, K. Tanikawa$^1$, S. Takahashi$^1$, and S. Maekawa$^{1,2}$}

\address{$^1$Institute for Materials Research, Tohoku University, Sendai
980-8577, Japan\\
$^2$CREST, Japan Science and Technology Agency (JST), Kawaguchi,
Saitama 332-0012, Japan }

\date{November 22, 2006}

\begin{abstract}
We study the two-impurity Anderson model for a semiconductor host
using the quantum Monte Carlo technique. We find that the impurity
spins exhibit ferromagnetic correlations with a range which can be
much more enhanced than in a half-filled metallic band. In
particular, the range is longest when the Fermi level is located
above the top of the valence band and decreases as the impurity
bound state becomes occupied. Comparisons with the photoemission
and optical absorption experiments suggest that this model
captures the basic electronic structure of Ga$_{1-x}$Mn$_x$As, the
prototypical dilute magnetic semiconductor (DMS). These numerical
results might also be useful for synthesizing DMS or dilute-oxide
ferromagnets with higher Curie temperatures.
\end{abstract}

\pacs{75.50.Pp, 75.30.Hx, 75.40.Mg, 71.55.-i}

\maketitle

The discovery of ferromagnetism in alloys of III-V semiconductors
with Mn started an intense research activity in the field of
dilute magnetic semiconductors (DMS) \cite{Maekawa,Ohno,Zutic}.
Room-temperature ferromagnetism in DMS would be a significant
development for spintronics device applications. In this respect,
it is important to understand the nature of the correlations which
develop between magnetic impurities in semiconductors and how they
differ from that in a metallic host. With this purpose, we present
exact numerical results on the two-impurity Anderson model for a
semiconductor host.

In order to study the multiple charge states of Au impurities in
Ge, the single-impurity Anderson model of a metallic host was
extended to the case of a semiconductor host using the
Hartree-Fock (HF) approximation \cite{Haldane}. After the
discovery of DMS, the magnetic properties of this model were
addressed within HF \cite{Ichimura,Takahashi}. It was shown that
long-range ferromagnetic (FM) correlations develop between
Anderson impurities in a semiconductor when the Fermi level is
located between the top of the valence band and the impurity bound
state (IBS), as illustrated in Fig. 1. The FM interaction between
the impurities is mediated by the impurity-induced polarization of
the valence electron spins, which are antiferromagnetically
coupled to the impurity moments. Within HF, the impurity-host
hybridization also induces host split-off states at the same
energy as the IBS. When the split-off state becomes occupied, the
spin polarizations of the valence band and of the split-off state
cancel. This causes the long-range FM correlations between the
impurities to vanish. Within the context of DMS, the Anderson
Hamiltonian for a semiconductor host was also considered by
Krstaji\'c {\it et al.} \cite{Krstajic}, and it was shown that an
FM interaction is generated between the impurities due to
kinematic exchange. In addition, this model was studied within HF
for investigating the multiple charge and spin states of
transition-metal atoms in hemoprotein \cite{Yamauchi}.

In this paper, we present quantum Monte Carlo (QMC) data on the
two-impurity Anderson model for a semiconductor host and make
comparisons with the HF results. We find that in a semiconductor
the nature of the magnetic correlations between the impurities is
different than in a metallic host. In particular, the impurities
exhibit long-range FM correlations when the Fermi level is located
above the top of the valence band, and the FM correlations weaken
as the IBS becomes occupied in agreement with HF
\cite{Ichimura,Takahashi}. Comparisons with the photoemission and
optical absorption experiments suggest that this model captures
the basic electronic structure of Ga$_{1-x}$Mn$_x$As. These
numerical results outline the parameter regime which yields the
longest-range FM correlations, and this information might be
useful for synthesizing DMS or dilute-oxide ferromagnets with
higher Curie temperatures.

The two-impurity Anderson model for a semiconductor host is
defined by
\begin{eqnarray} H && = \sum_{{\bf k},\alpha,\sigma}
(\varepsilon^{\alpha}_{{\bf k}}-\mu) c^{\dagger}_{{\bf
k}\alpha\sigma} c_{{\bf k}\alpha\sigma} + \sum_{{\bf
k},i,\alpha,\sigma} (V_{{\bf k}i} c^{\dagger}_{{\bf
k}\alpha\sigma} d_{i\sigma} \nonumber \\ && + {\rm H.c.}) +
E_d\sum_{i,\sigma} d^{\dagger}_{i\sigma} d_{i\sigma} + U\sum_i
n_{id\uparrow}n_{id\downarrow},
\end{eqnarray}
where $c^{\dagger}_{{\bf k}\alpha\sigma}$ ($c_{{\bf
k}\alpha\sigma}$) creates (annihilates) a host electron with
wavevector ${\bf k}$ and spin $\sigma$ in the valence ($\alpha=v$)
or conduction ($\alpha=c$) band, $d^{\dagger}_{i\sigma}$
($d_{i\sigma}$) is the creation (annihilation) operator for a
localized electron at impurity site $i$, and
$n_{id\sigma}=d^{\dagger}_{i\sigma} d_{i\sigma}$. The
hybridization matrix element is $V_{{\bf k}j} = V \exp(i {\bf
k}\cdot {\bf R}_j)$, where ${\bf R}_j$ is the coordinate of the
impurity site $j$. As usual, $E_d$ is the $d$-level energy, $U$ is
the onsite Coulomb repulsion, and $\mu$ the chemical potential.
The valence and the conduction bands have the forms
$\varepsilon^v_{{\bf k}} = - D \left( k / k_0 \right)^2$ and
$\varepsilon^c_{{\bf k}} = D \left( k / k_0 \right)^2 + \Delta_G$,
respectively (Figure 1). Here, $D$ is the bandwidth, $k_0$ is the
maximum wavevector and $\Delta_G$ is the semiconductor gap. In
this paper, we consider a two-dimensional semiconductor host with
a constant density of states $\rho_0=k_0^2/(4\pi D)$. We find
similar results for the three-dimensional case \cite{Tanikawa}. We
determine the energy scale by setting $D=12.0$. In addition, we
use $U=4.0$ and $E_d=\mu-U/2$ (the symmetric case), so that the
impurity sites develop large moments. We note that this
Hamiltonian is particle-hole symmetric with respect to
half-filling at $\mu=\Delta_G/2$. For the DMS materials, it is
estimated that $\Delta_G/D \approx 0.1$ to 0.2. We report results
for $\Delta_G=2.0$, values of the hybridization parameter
$\Delta\equiv \pi \rho_0 V^2$ ranging from 1.0 to 4.0, and inverse
temperature $\beta\equiv 1/T$ from 4 to 32. In order to study the
evolution of the magnetic correlations as we go from a metallic to
a semiconductor host, we will present results for $\mu$ from
$-D/2$ to $\Delta_G/2$.
\begin{figure}[t]
\centering
\includegraphics[width=4.5cm,bbllx=209,bblly=255,bburx=440,bbury=540,clip]{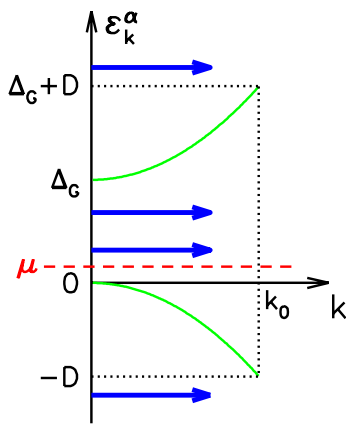}
\caption{(color online) Schematic drawing of the semiconductor
host bands $\varepsilon_k^{\alpha}$ (solid curves) and the
impurity bound states (thick arrows) obtained within HF. The
dashed line denotes the chemical potential $\mu$. } \label{fig1}
\end{figure}

The numerical results presented here were obtained with the
Hirsch-Fye quantum Monte Carlo technique \cite{Hirsch}. In the
following, we will first present results on the impurity
equal-time magnetic correlation function $\langle M_1^z
M_2^z\rangle$, where $M_i^z = n_{id\uparrow} - n_{id\downarrow}$
is the impurity magnetization operator. Next, we will present
results on the impurity single-particle spectral weight
$A(\omega)=-(1/\pi) {\rm Im} \,G^{\sigma}_{ii} (\omega)$, which is
obtained with the maximum-entropy analytic continuation technique
\cite{Linden} from the QMC data on the impurity Green's function
\begin{equation}
G^{\sigma}_{ii}(\tau) = - \langle T_{\tau}\, d_{i\sigma}(\tau)
d^{\dagger}_{i\sigma}(0)\rangle.
\end{equation}
Here, $T_{\tau}$ is the Matsubara time-ordering operator and
$d_{i\sigma}(\tau) = e^{H\tau} d_{i\sigma} e^{-H\tau}$. Since the
maximum-entropy procedure requires QMC data with very good
statistics, our results on $A(\omega)$ will be limited to the
high-temperature $\beta=8$ case. For lower $T$, we will discuss
QMC results on the impurity occupation number $\langle n_d\rangle
= \langle n_{id\uparrow} \rangle + \langle n_{id\downarrow}
\rangle$. We will also show data on the zero-frequency
inter-impurity magnetic susceptibility defined by
\begin{equation}
\chi_{12}(\omega=0) = \int_0^{\beta} d\tau \langle M_1(\tau)
M_2(0) \rangle.
\end{equation}
The following results were obtained using Matsubara time steps
$\Delta\tau=0.125$ and 0.25.

\begin{figure}[t]
\centering
\includegraphics[width=6.2cm,bbllx=95,bblly=140,bburx=525,bbury=715,clip]{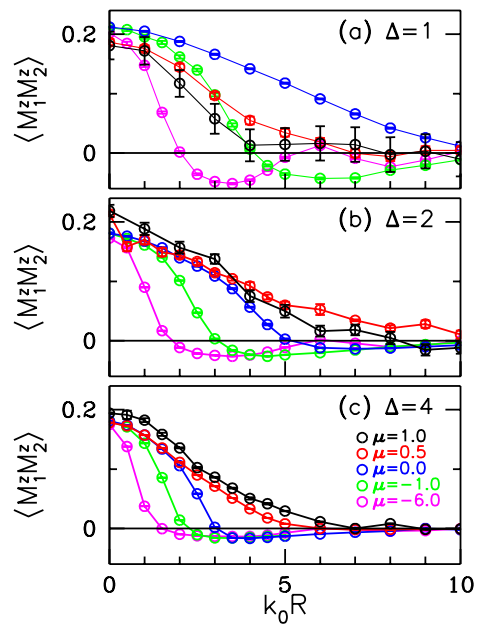}
\caption{(color online) $\langle M_1^z M_2^z\rangle$ vs $k_0R$
plotted at $\beta=16$ and various $\mu$ for hybridization (a)
$\Delta=1.0$, (b) 2.0, and (c) 4.0. } \label{fig2}
\end{figure}
Figures 2(a)-(c) show the impurity magnetic correlation function
$\langle M_1^z M_2^z\rangle$ vs $k_0R$, where $R=|{\bf R}_1-{\bf
R}_2|$ is the impurity separation, at $\beta=16$ for $\mu$ varying
from -6.0 to 1.0. Fig. 2(a) is for hybridization $\Delta=1.0$. At
$\mu=-6.0$, we observe oscillations in the $R$ dependence due to
an RKKY-type effective interaction between the impurities. These
results are similar to what has been obtained previously with QMC
for a half-filled metallic band \cite{Hirsch,Fye2,Fye}. The
wavelength of the oscillations increases when $\mu$ moves to
$-1.0$, because of the shortening of the Fermi wavevector. When
$\mu=0.0$, the impurity spins exhibit long-range FM correlations
at this temperature. We observe that upon further increasing $\mu$
to 0.5 or 1.0, the FM correlations become weaker. This is because
the IBS becomes occupied as $\mu$ changes from 0.0 to 0.5, as will
be seen in Fig. 3(a). In Figs. 2(b) and (c), results on $\langle
M_1^z M_2^z \rangle$ are shown for $\Delta=2.0$ and 4.0,
respectively. In Fig. 2(b), we observe that $\langle M_1^z
M_2^z\rangle$ has the slowest decay for $\mu=0.5$, while in Fig.
2(c) this occurs for $\mu=1.0$. We find that the impurity
occupation $\langle n_d\rangle$ increases between $\mu=0.5$ and
1.0 for $\Delta=2.0$ and $\beta=16$. In addition, for $\Delta=4.0$
and $\beta=8$, the maximum-entropy image of $A(\omega)$ shows that
the IBS is located at $\omega\approx 1.0$. Hence, we observe that
the range of the FM correlations for the semiconductor is
determined by the occupation of the IBS in agreement with the HF
predictions \cite{Ichimura,Takahashi}. In Figs. 2(a)-(c), it is
also seen that the range increases with decreasing $\Delta$.

\begin{figure}[t]
\centering
\includegraphics[width=6cm,bbllx=96,bblly=212,bburx=525,bbury=494,clip]{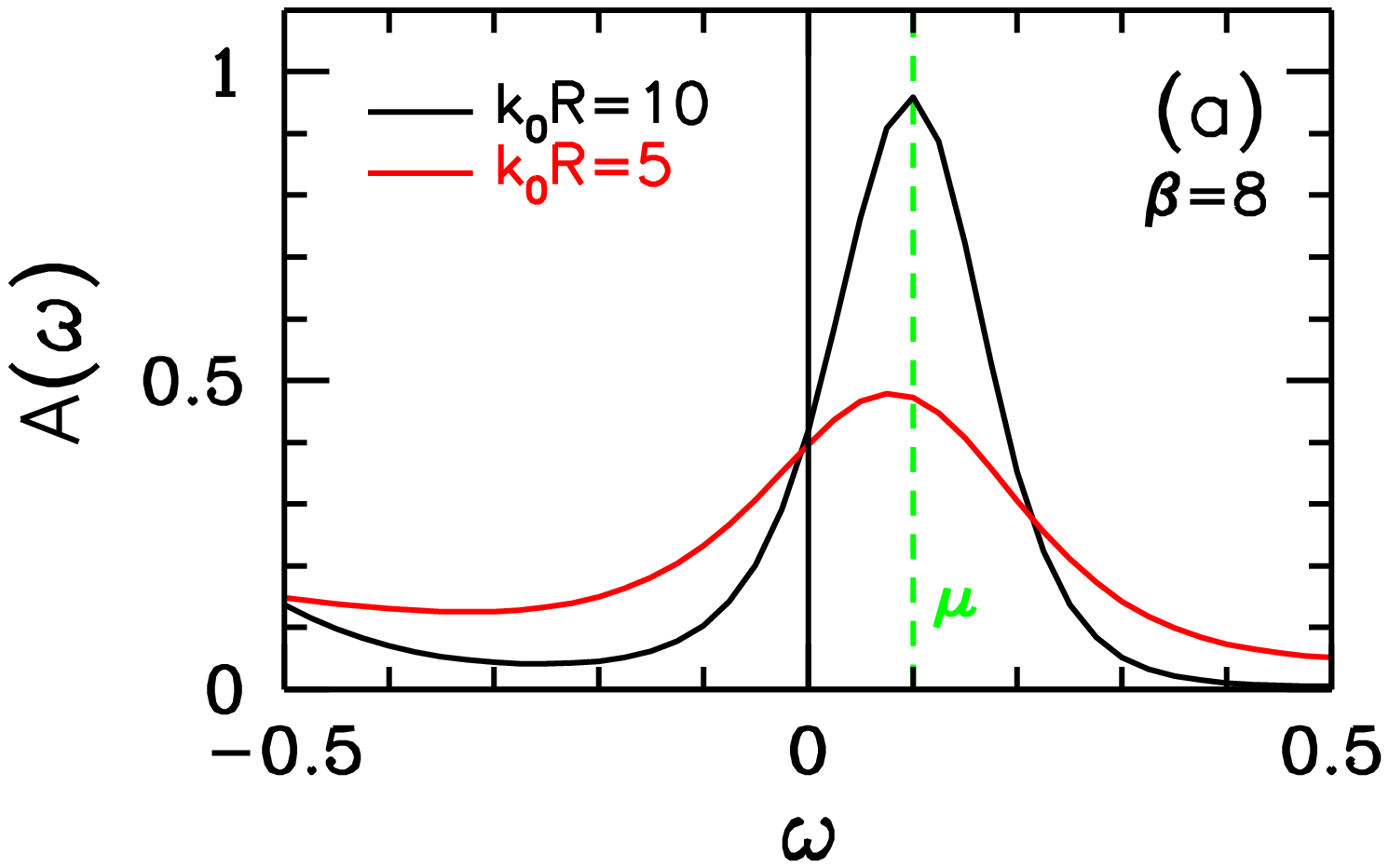}\\
\includegraphics[width=6cm,bbllx=96,bblly=212,bburx=525,bbury=707,clip]{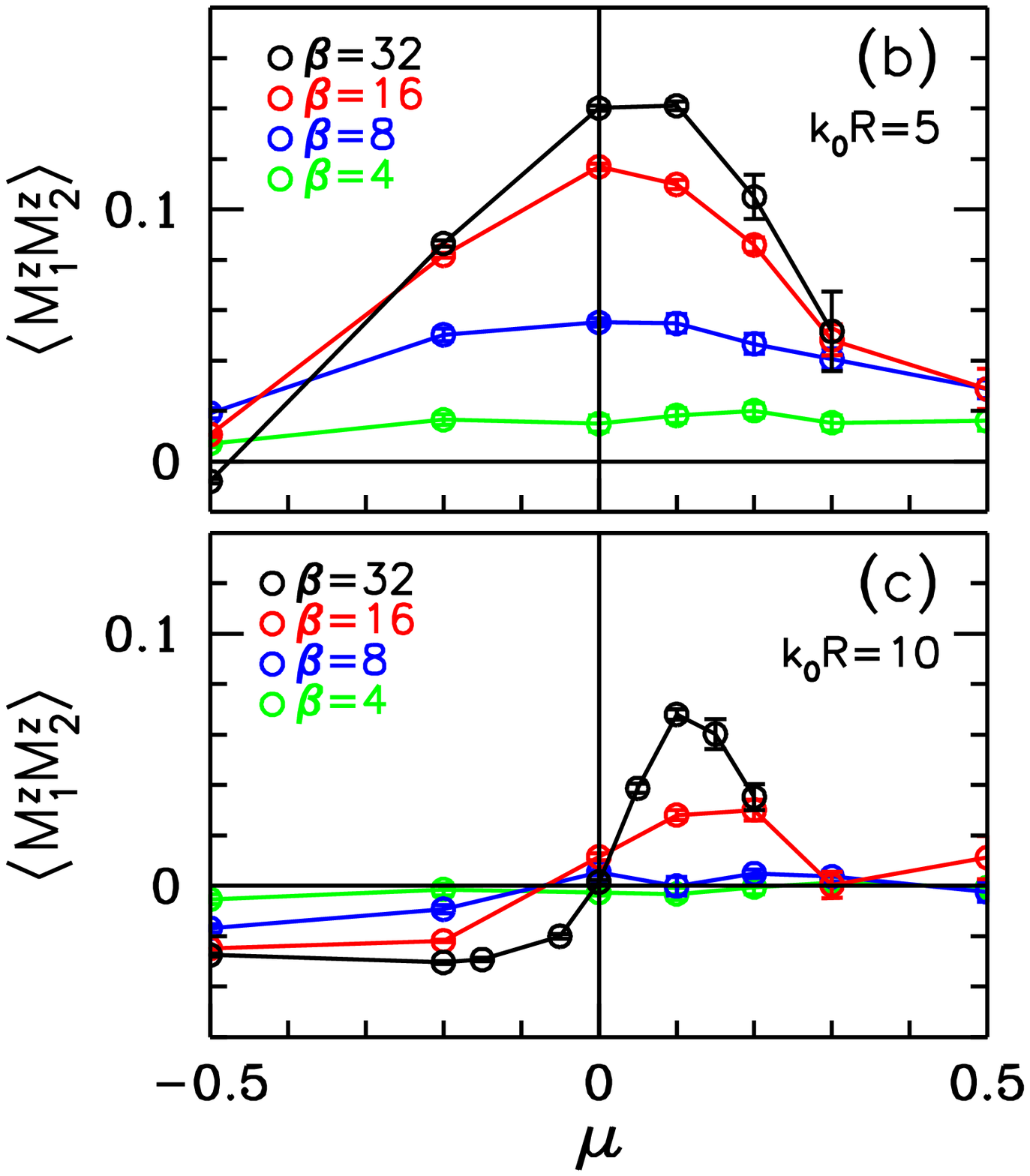}
\caption{(color online) (a) Impurity single-particle spectral
weight $A(\omega)$ vs $\omega$ for $k_0R=5$ and 10 at $\beta=8$.
Here, the vertical dashed line denotes $\mu$, and the top of the
valence band is located at $\omega=0$. In (b) and (c), $\langle
M_1^z M_2^z\rangle$ vs $\mu$ is plotted for $k_0R=5$ and 10,
respectively, at various $\beta$. These results are for
$\Delta=1.0$. } \label{fig3}
\end{figure}
In Figs. 3 and 4, we discuss the $\Delta=1.0$ case in more detail.
In Fig. 3(a), the impurity spectral weight $A(\omega)$ vs $\omega$
is plotted for $\beta=8$, $\mu=0.1$, and $k_0R=5$ and 10. Here,
the $\omega$-axis has been shifted so that the top of the valence
band is located at $\omega=0$. For $k_0R=10$, we observe a peak at
$\omega_{BS}\approx 0.1$ in the semiconductor gap, which we
identify as the IBS. For $k_0R=5$, the bound state is broader due
to stronger correlations between the impurities. However, we also
find that $A(\omega)$ exhibits significant $T$ dependence at
$\beta=8$, and Fig. 3(a) does not yet represent the low-$T$ limit.
Next, in Figs. 3(b) and (c), $\langle M_1^z M_2^z \rangle$
evaluated at $k_0R=5$ and 10 is plotted as a function of $\mu$.
Fig. 3(b) shows that, at low $T$ for $k_0R=5$, $\langle M_1^z
M_2^z \rangle$ decreases when $\mu \gtrsim 0.25$. For this value
of $k_0R$ and $\beta=32$, we find that the impurity occupation
$\langle n_d\rangle$ develops a step discontinuity at $\mu \approx
0.25$, which is consistent with the decrease of $\langle M_1^z
M_2^z\rangle$ when  $\mu\gtrsim 0.25$. For $k_0R=10$ and
$\beta=32$, both $\langle M_1^z M_2^z \rangle$ and $\langle n_d
\rangle$ exhibit significant $T$ dependence in the vicinity of the
semiconductor gap edge. These results show that $\langle M_1^z
M_2^z\rangle$ depends strongly on the value of $\mu$.

\begin{figure}[t]
\centering
\includegraphics[width=6cm,bbllx=96,bblly=212,bburx=525,bbury=494,clip]{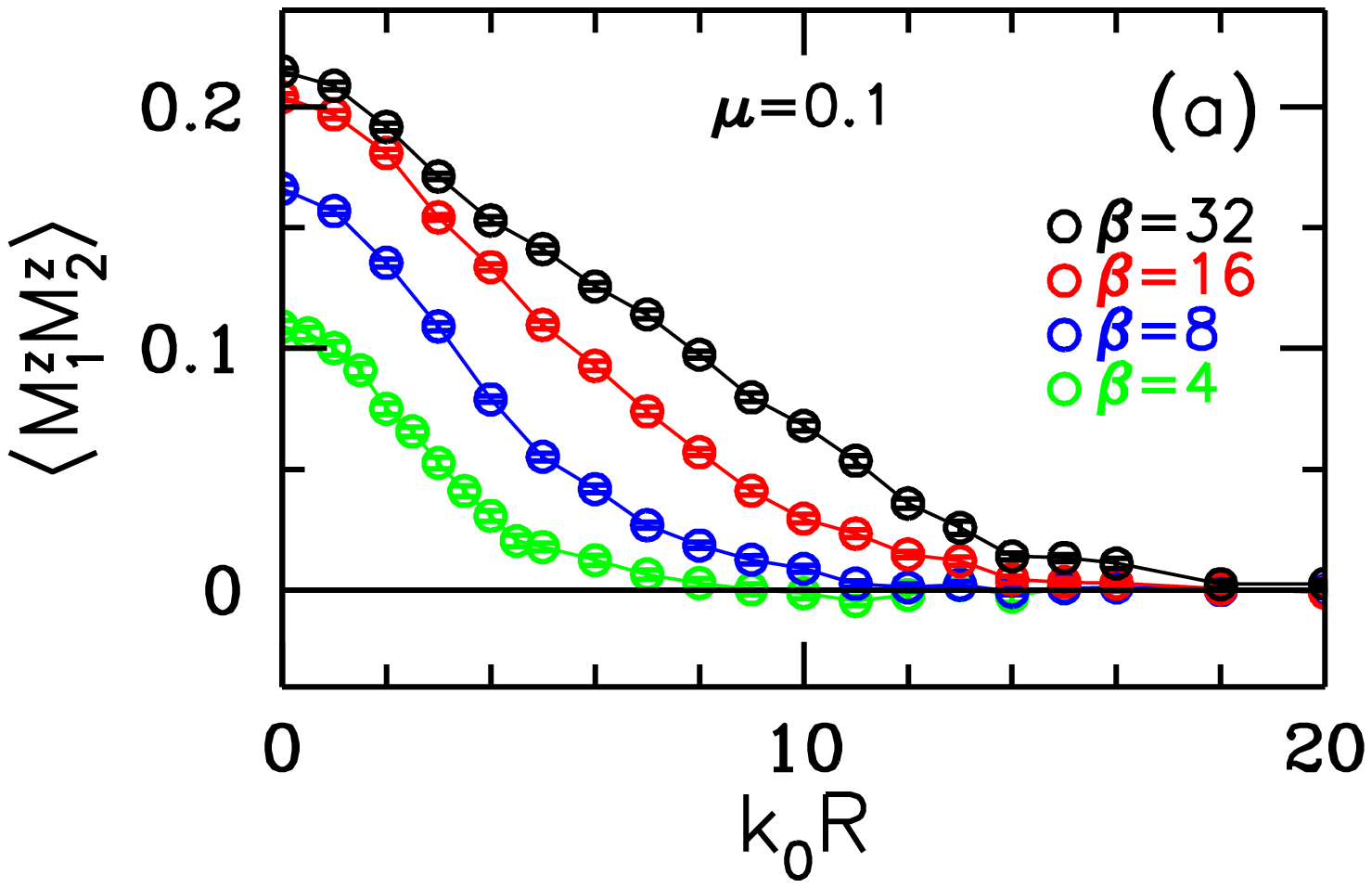}\\
\includegraphics[width=6cm,bbllx=96,bblly=212,bburx=525,bbury=494,clip]{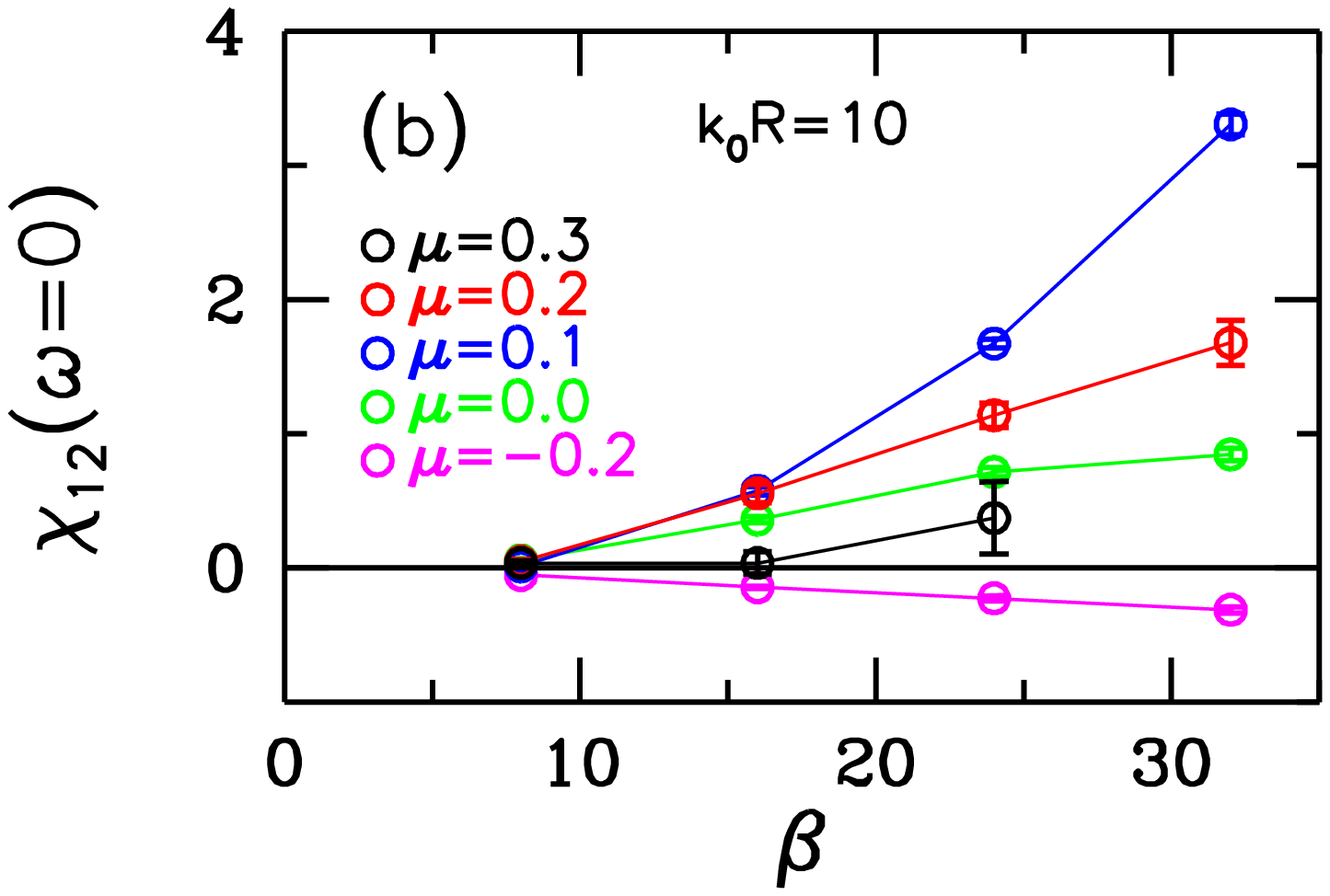}
\caption{(color online) (a) $\langle M_1^z M_2^z\rangle$ vs $k_0R$
for $\mu=0.1$ at various $\beta$. (b) Inter-impurity magnetic
susceptibility $\chi_{12}(\omega=0)$ vs $\beta$ for $k_0R=10$ at
various $\mu$. These results are for $\Delta=1.0$. } \label{fig4}
\end{figure}
Figure 4(a) shows the temperature dependence of $\langle M_1^z
M_2^z \rangle$ vs $k_0R$ for $\mu=0.1$. We observe that, at
$\beta=32$, the range of the FM correlations is enhanced by about
an order of magnitude with respect to that in a half-filled
metallic band. In Fig. 4(b), the $T$ dependence of the
inter-impurity susceptibility $\chi_{12}(\omega=0)$ is shown for
$k_0R=10$ and various values of $\mu$. We observe that, as $\beta$
increases, $\chi_{12}$ becomes strongly enhanced for $\mu =0.1$,
while it remains weak for $\mu=0.3$. At $\mu=-0.2$, $\chi_{12}$ is
antiferromagnetic for this value of $k_0R$. Figures 3 and 4 show
that long-range FM correlations develop between the impurities
depending on the position of $\mu$.

In the QMC simulations, we find that $\langle M_1^z M_2^z \rangle$
has larger error bars when IBS is occupied. This might be due to
the cancellation of the host spin polarizations originating from
the split-off state and from the valence band. The intra-impurity
measurements such as $\langle (M_1^z)^2 \rangle$ or
$G_{ii}^{\sigma}(\tau)$ do not exhibit such behavior.

Within HF, the range $\xi$ of the FM correlations between the
impurities is determined by the energy of the IBS and, hence, the
energy of the split-off state: $\xi \approx (16\pi \rho_0
\omega_{BS})^{-1/2}$ for a constant density of states and
semi-infinite host bands. In particular, the spatial extent of the
spin polarization of the valence band around the impurity is given
by $4\rho_0 \Delta Z_d e^{-r/\xi}/(r/\xi)$, where $Z_d$ is the
weight of the pole at $\omega_{BS}$ in the impurity
single-particle Green's function. Within HF, $\omega_{BS}$
decreases rapidly as $\Delta/\Delta_G \rightarrow 0$, which agrees
with the $\Delta$ dependence of the range seen in Figs. 2(a)-(c).
In addition, for $T=0$, $\Delta=1.0$, and $0< \mu < \omega_{BS}$,
HF yields $\omega_{BS}\approx 0.02$ and $k_0\xi \approx 9$. We
note that the effects of the inter-impurity correlations on the
impurity single-particle Green's function are neglected within
this approximation.

In the QMC and HF calculations, the location of the Fermi level
with respect to the IBS is important; the FM correlations weaken
as the IBS becomes occupied. Photoemission and optical
measurements on Ga$_{1-x}$Mn$_x$As provide evidence that the
occupation of the Mn-induced impurity band is similarly important
for the magnetic properties of this prototypical DMS ferromagnet.
Photoemission experiments \cite{Okabayashi} observed an Mn-induced
state above the valence band and right below the Fermi level in
Ga$_{1-x}$Mn$_x$As. Clearly, inverse photoemission experiments are
required to detect the unoccupied portion of the Mn-induced
impurity band. STM experiments also observed the impurity band in
this compound \cite{Yakunin}. Recent optical-absorption
measurements \cite{Burch}, which show a redshift of the
mid-infrared peak with Mn doping, provide evidence that the Fermi
level is located in the Mn-induced impurity band in
Ga$_{1-x}$Mn$_x$As. Furthermore, as the impurity band becomes less
occupied with Mn doping, the Curie temperature $T_c$ increases in
annealed samples, which is in agreement with the QMC and HF
results. The comparisons of these experiments and the numerical
results suggest that the Anderson Hamiltonian for a semiconductor
host provides a basic electronic model for the DMS ferromagnets.

These numerical results also suggest that, in addition to Mn
substitution, a possible way of increasing $T_c$ in
Ga$_{1-x}$Mn$_x$As is to decrease the impurity-band occupation by
changing the semiconductor host material or by using additional
dopants. Alternative ways of enhancing the FM correlations in this
model is provided by varying the hybridization parameter $\Delta$
or the semiconductor gap $\Delta_G$. The QMC simulations show that
$\xi$ increases as $\Delta$ goes from 4.0 to 1.0. Within HF, $\xi$
can take very large values as $\Delta/\Delta_G$ decreases.
Recently, $T_c$'s exceeding the room temperature have been
reported in dilute oxides such as ZnO and TiO$_2$ with transition
metal impurities \cite{Matsumoto,Sharma}. At this point, it is
important to determine whether the Anderson model of a magnetic
impurity applies to this case. Obviously, experiments probing the
electronic state of the FM dilute oxides are necessary to answer
this question.

The two-impurity Anderson model for a semiconductor host might be
oversimplified for describing the ferromagnetism of the DMS. Our
calculations are for spin-1/2 Anderson impurities, and we have
neglected the multi-orbital structure of the spin-5/2 Mn
impurities. Hence, the effects of the atomic Hund's rule couplings
are not included. In addition, we neglect the long-range Coulomb
repulsion between the impurity and the host electrons. Keeping
these caveats in mind, it is interesting to note that the
theoretical studies of high-$T_c$ DMS ferromagnetism could have
preceded the experimental discovery, if the Hirsch-Fye algorithm
had been applied to a semiconductor host when it was developed
twenty years ago.

In summary, we have presented QMC results to show that long-range
FM correlations develop between magnetic impurities in
semiconductors. In particular, the FM correlations have the
longest range when the Fermi level is located above the top of the
valence band, and they weaken as the IBS becomes occupied. Hence,
the position of the Fermi level with respect to the IBS plays a
crucial role in determining the range of the FM correlations in
agreement with HF. Comparisons with the photoemission and optical
absorption experiments suggest that the two-impurity Anderson
model in a semiconductor host captures the basic electronic
structure of Ga$_{1-x}$Mn$_x$As. The numerical results presented
here outline the parameter regime which yields the longest-range
FM correlations, and this information might be useful for
synthesizing higher-$T_c$ DMS materials.

We thank V.A. Ivanov for bringing Ref. \cite{Krstajic} to our
attention and for useful comments. This work was supported by the
NAREGI Nanoscience Project and a Grant-in Aid for Scientific
Research from the Ministry of Education, Culture, Sports, Science
and Technology of Japan, and NEDO. One of us (N.B.) gratefully
acknowledges support from the Japan Society for the Promotion of
Science and the Turkish Academy of Sciences
(EA-TUBA-GEBIP/2001-1-1).


\end{document}